\documentclass[a4paper, 12pt]{article}

\usepackage{relsize}
\usepackage{times,float}
\usepackage{amsmath}
\usepackage{amsthm}
\theoremstyle{definition}
\usepackage{amssymb}
\usepackage{graphicx,epsf}
\usepackage{color}

\usepackage{authblk}

\usepackage{bm}
\usepackage{cleveref}
\usepackage[subnum]{cases}
\usepackage[english]{babel}
\usepackage[titletoc,title]{appendix}
\usepackage{fancyhdr}
\usepackage{lscape}
\usepackage{caption}
\usepackage{subcaption}
\usepackage{enumerate}
\usepackage{booktabs}
\usepackage{multirow}
\usepackage{graphicx}
\usepackage{float}
\usepackage{epstopdf}
\usepackage{multirow,array}
\usepackage{array}
\usepackage{fancyvrb}
\usepackage[english]{babel}
\usepackage[round]{natbib}   
\setcitestyle{authoryear,open={(},close={)}}
\usepackage{csquotes}
\usepackage{tikz}
\usetikzlibrary{backgrounds,automata}
\usetikzlibrary{patterns}
\usetikzlibrary{positioning}
\usetikzlibrary{arrows,positioning,shapes.geometric}
\usetikzlibrary{decorations.markings}

\usepackage{tikz}

\usepackage{graphicx}

\renewcommand{\mkbegdispquote}[2]{\itshape}

\newcolumntype{C}{>{\centering\arraybackslash}p{2ex}}

\usepackage{ bbold }

\usepackage{graphicx}
\usepackage{tikz}

\def\ignore#1{}
\newlength{\defbaselineskip} 
\newcommand{\setlinespacing}[1]%
{\setlength{\baselineskip}{#1 \defbaselineskip}}

\makeatletter

\newcommand*{\Rom}[1]{\expandafter\@slowromancap\romannumeral #1@}
\makeatother
\crefname{proposition}{Proposition}{Proposition}
\crefname{assumption}{Assumption}{Assumption}
\crefname{theorem}{Theorem}{Theorem}
\crefname{corollary}{Corollary}{Corollary}
\crefname{example}{Example}{Example}
\crefname{lemma}{Lemma}{Lemma}
\crefname{remark}{Remark}{Remark}
\crefname{section}{Section}{Section}
\crefname{definition}{Definition}{Definition}
\crefname{note}{Note}{Note}
\crefname{condition}{Condition}{Condition}

\newtheorem{lemma}{Lemma}
\newtheorem*{lemma*}{Lemma}
\newtheorem{definition}{Definition}
\newtheorem{theorem}{Theorem}
\newtheorem{proposition}{Proposition}
\newtheorem*{proposition*}{Proposition}
\newtheorem*{theorem*}{Theorem}

\newtheorem*{proof*}{Proof}
\newtheorem*{definition*}{Definition}
\newtheorem{example}{Example}
\newtheorem*{example*}{Example}
\newtheorem{remark}{Remark}
\newtheorem*{remark*}{Remark}

\newtheorem*{assumption*}{Assumption}

\newtheorem*{condition*}{Condition}

\newtheorem{corollary}{Corollary}

\usepackage{comment}

\usetikzlibrary{decorations.pathreplacing,angles,quotes}

\begin{document}
\title{{\textbf{Information Sale on Network}}\thanks{Thanks to seminar participants in UHM, FDU, DKU for useful comments. All errors are our own. Financial Support from the Ministry of Education of China Grant No.19YJC790033 are greatly appreciated.}}
\author[1]{Jihwan Do \thanks{e-mail: econ.jihwando@yonsei.ac.kr}}
\author[2]{Lining Han \thanks{e-mail: hanlining@whu.edu.cn}}
\author[3]{Xiaoxi Li \thanks{e-mail: xli\_whu@163.com}}
\affil[1]{School of Economics, Yonsei University}
\affil[2,3]{Economics and Management School, Wuhan University}

\maketitle

	\vspace{-1em}
\begin{center} 
This is a preliminary version. Please do not distribute. 
\end{center}

\begin{abstract}
This paper studies a stylized model of a monopoly data seller when \emph{information-sharing network} exists among data buyers. We show that, if the buyers' prior information is sufficiently noisy, the optimal selling strategy is characterized by a \emph{maximum independent set}, which is the largest set of buyers who do not have information-sharing link at all. In addition, the precision of the seller's data decreases in the number of information-sharing links among buyers, but it is \emph{higher} than the socially efficient level of precision. 
Based on the results, we also examine profit-improving intervention strategies such as link removals and node isolation. 
\end{abstract}
	\vspace{1em}
	

\clearpage
\setcounter{page}{1}

\section{Introduction}  \label{sec:intro}

The business of collecting and selling data has become a growing industry that spans various sectors, including finance, healthcare, retail, and marketing. The companies that collect data can monetize it by selling it to interested third parties, and the advance in information technologies allows transactions of data in various formats, ranging from raw data feeds to customized research reports. The demand for data has also grown exponentially in recent years thanks to the rise of artificial intelligence, machine learning, and data analytics, which have made it easier to extract insights from data. Platforms such as WRDS, Bloomberg, and Datastream, for example, make significant profits by selling access to various financial, economic, and social science data to academic institutions and corporate agents.

Unsurprisingly, the value of data could be significantly influenced by the social networks among potential data buyers.  For instance, researchers typically share their research by posting updates on their personal websites or presenting at conferences in local communities; it is also common for employees of one institute to share the access right of a prescribed date set privately with their peers in a different institute. In this case, an agent with a bigger social network may have a lower willingness to pay for additional data in that s/he may learn from peers, and such a social network among data buyers may limit the ability of a data seller to increase its profit. Therefore, the optimal selling strategies of the data seller must take into account the existing data-sharing networks among buyers, which, in turn, affects the market outcomes such as efficiency and welfare.

Motivated by these observations, this paper studies the role of the information-sharing network among data buyers in characterizing the optimal selling strategy for a monopolistic data vendor and the resulting data prices, qualities, and social and consumer welfare. 
To this end, we develop a stylized quadratic-Gaussian payoff model in which the monopoly data seller first chooses the quality (i.e., precision of informative signals) and prices of data, and then the agents on the social network simultaneously choose whether to buy the data. A key assumption---which we believe is a realistic description of many data industries---is that the seller may offer different prices to different buyers (i.e., price discrimination is feasible), but is not allowed to generate or offer different data qualities (i.e., quality discrimination is not feasible). As a result, the monopoly seller's problem becomes to finding an optimal \emph{set of buyers} who are led to buying the data and the data quality that these buyers will enjoy.

In this environment, we show that, when the buyers' prior information is sufficiently noisy, the optimal selling strategy is characterized by a \emph{maximum independent set}, which is the largest set of buyers who do not have information-sharing link at all. This result greatly simplifies further analysis in that the optimal data quality and prices, as well as the seller's profits, depend solely on the number of buyers in a maximum independent set (whose cardinality is independent of the selection of such a set) rather than the entire topology of the social network. In particular, the equilibrium data quality and the seller's profit are shown to be increasing in the number of agents in a maximum independent set. Compared to the socially efficient level, however, the precision of data that is optimal for the seller is always higher, meaning that \emph{too precise information} is provided to each individual buyer from a social welfare perspective.  The consumer welfare, on the other hand, turns out to depend on more details of the network and which maximum independent set is chosen in equilibrium, and illustrate this observation in several examples. We also examine profit-improving intervention strategies such as link removals or node isolation and provide a characterization of Pareto-efficient networks.

\subsection{Related Literature} \label{sec:lit}

This paper is related to four streams of existing literature. Firstly, the decisions of the data consumers on a network are related to the research on strategic interactions of players on a network, which is widely explored. 
For instance, \cite{ballester2006} study the games of strategic complements on the network. \cite{bramoulle2007public} study the games of strategic substitutes on the network. 
\cite{bramoulle_strategic_2014} provide a general, unifying analysis of the strategic interaction of players on an exogenous network, the paper studies the existence, uniqueness, and stability of equilibrium. 

Secondly, consumers seek to take actions close to some unknown state of the world as a fundamental motive is studied in the research of costly information acquisition. 
\cite{myatt2012endogenous} study coordination game in which players choose the level of the cost paid to acquire information from various sources, and characterize the unique linear equilibrium. 
\cite{myatt2018information} study the price-setting problem of oligopolies, who have access to various information sources, and discuss the information use and acquisition of oligopolies of different sizes. 
\cite{yang2015coordination} study the flexible information acquisition in a coordination game. 
\cite{grossman1980impossibility} study the incentives of individuals to acquire information with cost and implications for capital markets. 
\cite{page2017experimental} design an experiment to study the information acquisition in experimental asset markets, where the traders can acquire information costly, and find a phenomenon of over-acquisition. \cite{bloch2018rumors} study the transmission of rumors in social networks and show that social networks can act as a filter.

Thirdly, a stream of recent research studies the problem of information acquisition through networks. 
\cite{galeotti2010law} develop a model where players acquire information and form endogenous information access with others, they find that the robust equilibrium satisfies the law of the few.  
\cite{goyal2017information} conduct an experiment about information acquisition and linking, and they study the effect of the relative cost of information acquisition and linking on the equilibrium and welfare. 
\cite{leister2020information} study the information acquisition and use in network games of both strategic complements and substitutes. The paper analyzes the welfare of equilibrium and discusses policy intervention. 
\cite{myatt2019} study information acquisition and use in network games of strategic complements, where the players receive signals from multiple sources costly. They analyze the equilibrium with asymmetric players.  
\cite{herskovic2020} develop an endogenous network formation model, in which players form links to acquire information and then play a game of strategic complements based on the information. The paper finds that the equilibrium network is core-periphery. 
\cite{murayama2019social} study the information used in network and social value of public information. 
There is also research about information acquisition on the network in the asset pricing model. 
\cite{ozsoylev2011asset} study the information acquisition problem of asset pricing model in a large network. They consider a general network with sparse connections and find the rational expectation equilibrium exists in a sparse network and connection proportional to a number of agents.   
\cite{han2013social} study the information acquisition on the social networks in the asset pricing model. They consider the island-connection network and find that more connections of the network improve market efficiency, and increase liquidity given the total amount of information, but reduce the incentive of acquiring information. 
Besides the theoretical research about information acquisition on networks, \cite{halim2019costly} design an experiment to study the incentive of acquiring information in social networks.  

Moreover, the information transmission among players is also related to research about information sharing on the supply chain. \cite{gal1985information}, \cite{li1985cournot} study the incentive of sharing private information in a horizontal market structure. A more recent literature studies information sharing in vertical market structures, such as supply chains. \cite{lee2000value} study the value of information sharing in the supply chain, and find information sharing mitigates the vertical information distortions and brings lower costs. \cite{li2002information} study the information sharing in the supply chain with competition among the retailers. 
\cite{liu2021information} study the incentive of the retail platforms to share information among multiple competing sellers.

Finally, the optimal selling strategy of the seller studied in this paper is related to the literature on optimal pricing and targeting on a network. \cite{Bloch2016} provides a good survey for targeting and seeding problems on networks. 
\cite{campbell2013word} studies advertising with word-of-mouth communication between friends and finds the target for advertising is not the individuals with the most friends.
\cite{Candogan2012} study the optimal pricing in networks with positive externalities of a monopolistic seller, and find that the monopolist offers a discount, which is proportional to the Bonacich centrality, to the central buyers. 
\cite{Bloch2013} study the optimal monopoly pricing in social networks where agents care about the consumption or prices of their neighbors. 
Our research is also related to papers about persuasion on networks, such as \cite{egorov2020persuasion} and \cite{candogan2022persuasion}. \cite{egorov2020persuasion} find that the optimal propaganda might target peripheral agents rather than central agents. \cite{cheng2022networks} find the networks that permit stable allocations depend on the size of the independent set, which is related to our findings of the optimal contract.

\section{Model} \label{sec:model}
We consider the simplest model of a monopoly data seller's problem who trades with finitely many data buyers, $N=\{1,2,...,n\}$.  
Each buyer $i$'s payoff is determined by his action $a_i\in\mathbb{R}$ and true state $\theta\in\mathbb{R}$ as follows:
$$u_{i}\left(a_i,\theta\right)=-\left(a_{i}-\theta\right)^{2},$$
where $\theta\sim N\left(0,\frac{1}{z_0}\right)$ with $z_0>0$. The buyers know the prior distribution of $\theta$. The data seller can sell a copy of the data to the buyer $i$, and the buyer $i$ can buy the data and learn a signal of the form,
$$s_{i}=\theta +\epsilon_{i},$$ 
where $\epsilon_{i} \sim N\left(0, \frac{1}{z_{i}}\right)$ is independent of $\theta$ and $\epsilon_j$ for all $j\neq i$. We will denote by $C_i=\left(z_{i},p_i\right)$ a contract offered to buyer $i$, where $z_{i} \geq 0$ measures the precision of the signal $s_i$ (i.e., data quality) and $p_i\geq 0$ is the price paid to the seller, respectively. If buyer $i$ does not buy access to the data, then he does not learn the signal $s_{i}$.
In the following, we consider homogenous buyers, who have the same learning ability and generate the signals with the same precision if they receive the same data. Thus, the accuracy of signal $z_{i}$ can be interpreted as the amount of data sold to buyer $i$. Moreover, the buyers have the same learning capacity that they could only learn one signal from the data.

Consider the case where the data seller can limit the buyer's access to the data, thus the buyer is not able to share the usage of the data with others. However, the buyer can share his interpretation of the data, which is his signal, with his friends. 
To capture the information spillover between buyers, we introduce an undirected social network $G$ of buyers as follows: $g_{ij}=g_{ji}=1$ if $i$ and $j$ are connected. Assume $g_{ii}=1$ for all $i$. The buyers $i$ and $j$ can observe each other's signals if and only if they are connected. The set of the neighbors of $i$ and its cardinality are denoted by $N_i$ and $n_i$, respectively.\footnote{Agent $i$ is not in the set of his neighbors, $i \notin N_i$.} The buyers $-i$ denotes the set of buyers $N \setminus \{i\}$.
\medskip
The timeline of the problem is summarized as follows: 
\begin{itemize}
	\item [(1)] The seller simultaneously offers the contracts $C=(C_1, C_2, \dots, C_n)$, which is publicly observed by the buyers.
	\item [(2)] Each buyer simultaneously and independently chooses to accept or reject the offered contract.
	\item [(3)] Each buyer observes the signals $\left(s_{j}:j\in N_i\right)$ of his neighbors and his own signal $s_i$ if he accepts his contract offered with positive precision.
	\item [(4)] Each buyer chooses an action $a_i$.
\end{itemize}
\medskip

Note that it is without loss of generality to assume that every buyer accepts a contract in equilibrium: a contract rejected by a buyer is equivalent to a null contract $(0,0)$ accepted by the buyer. Thus, the payoff of a seller given contracts $C=(z_i,p_i)_{i\in N}$ is 
$$\pi\left(C\right)=\sum_{i\in N}p_i-c\left(z_1,...,z_n\right),$$
where $c(z_1,...,z_n)$ is a cost incurred for the seller to collect and process data. On the other hand, the buyer $i$'s payoff from $C$ is 
$$\pi_{i}\left(a_i,C\right)=E\left(-\left(a_{i}-\theta\right)^{2}|\left(s_{i},s_{j}, z_i,z_j:j\in N_i\right)\right)-p_{i}.$$

\bigskip

Throughout the paper, we will impose the the following key assumption, which we believe is a realistic description of many data industries: the seller may offer different prices to different buyers (i.e., price discrimination is feasible), but is not allowed to generate or offer different data qualities (i.e., quality discrimination is not feasible). More precisely, this is equivalent to assuming $z_{i} = z > 0$ or $z_{i}=0$ for all $i$, so  $z_{i} > 0$ means that the seller offers the usage of the data in contract $C_{i}$, and the buyer generates a signal with precision $z$ after learning; whereas $z_{i}=0$ means that the seller offers a null contract to buyer $i$, so that buyer $i$ does not have access to the data and can not generate any signal with positive precision. Since the uniform $z$ measures the amount of data collected by the seller, we will often say that the seller collects data with precision $z$ and assume the constant marginal cost $c(z)=\gamma z$ with $\gamma>0.$

\section{Analysis} \label{sec:eq}
\subsection{Consumer Behaviors} \label{sec:consumer}
First, consider a buyer's optimal choice of action (when every buyer accepts the contract $C$):
$$\max_{a_i}E\left(-\left(a_{i}-\theta\right)^{2}|\left(s_{i},s_{j},z_i,z_j:j\in N_i\right)\right).$$ 

\begin{lemma} \label{lem:opt_act}
The optimal action of agent $i$ is a linear function of the observed signals as follows:   
$$a_{i}^*(z)=\sum_{j=1}^{n}\left(\frac{g_{i j} z_{j}}{z_0+\sum_{j=1}^{n} g_{i j} z_{j}}\right)s_{j}.$$ 
\end{lemma}

Given this, buyer $i$'s expected payoff when accepting $C_i$ is 
\begin{align*}
    E\left(-\left(a_{i}^*(z)-\theta\right)^{2}|\left(s_{i},s_{j},z_i,z_j:j\in N_i\right)\right)-p_i  =-\left(\frac{1}{z_0+\sum_{j=1}^{n} g_{i j} z_{j}}\right)-p_i.
\end{align*}
If the buyer $i$ rejects $C_{i}$, on the other hand, then the signal $s_i$ is no longer available but still $i$ can observe the signals of his neighbors. Therefore, the expected payoff when rejecting $C_i$ is given by
\begin{align*}
    E\left(-\left(a_{i}^*(z)-\theta\right)^{2}|\left(s_{j},z_j:j\in N_i\right)\right) =-\left(\frac{1}{z_0+\sum_{j\neq i} g_{i j} z_{j}}\right).
\end{align*}

The following lemma characterizes the maximal price that buyer $i$ is willing to pay for access to data and receive an extra signal.

\begin{lemma} \label{lem:contract}
Given the contracts $C$, buyer $i$ is willing to accept the contract $C_{i}$ if and only if
$$p_{i} \leq \frac{1}{\sum_{j \neq i} g_{i j} z_{j}}-\frac{1}{\sum_{j=1}^{n} g_{i j} z_{j}}=\frac{ z_i}{(\sum_{j\neq i} g_{i j}z_j)(\sum_{j=1}^{n} g_{i j}z_j)}.$$ 
\end{lemma}

Note that, when only one buyer exists, one can derive the optimal contract immediately from this lemma.

\begin{example} \label{ex:1b}
For the case $n=1$, the profit of the seller is $\pi(C) = \frac{1}{z_{0}} - \frac{1}{z_{0} + z_{1}} - \gamma z$. From the first-order condition, we have
$$z_{1}+z_0 = \frac{1}{\sqrt{\gamma}}$$
so, the data price is  
$$p_{1} = \frac{1}{z_{0}} - \frac{1}{z_{0} + z_{1}} = \frac{1}{z_0} - \sqrt{\gamma}$$
and the precision of signal is
$$z_{1} = \frac{1}{\sqrt{\gamma}}-z_0.$$ 
Thus, the optimal contract for the seller is 
$$C = \left(\frac{1}{\sqrt{\gamma}}-z_0, \frac{1}{z_0}-\sqrt{\gamma}\right).$$ 
\end{example}

As illustrated in this example, the buyer would have no incentive to buy access to data and learn a new signal if the common prior is accurate enough, and in this case, the contract offered by the seller becomes trivial. Indeed, this observation holds in general, and we will restrict our attention to the case of a sufficiently noisy common prior to guaranteeing that the optimal data quality is strictly positive.

\subsection{Optimal Contract}\label{sec:main}

Let $G$ be an undirected network with $N$ the set of users. Given the marginal cost of collecting data $\gamma$, and the precision of common prior. The seller offers a contract $C=(z_{i}, p_{i})_{i \in N}$ to the buyers on network $G$ to maximize his payoff $\pi(C)$. For the problem of selling an indivisible database to homogenous buyers on a network, the seller selects a subset of buyers as the target to sell access to the data. $M(C)$ denotes the target of sale, there is 
$$M(C) =\{i \in N | z_{i}>0, C=(z_{i}, p_{i})_{i \in N}\}.$$ 

In the following analysis, we introduce some notations. Denote $N_i=\{j\in N|g_{ij}=1\}$ as the set of buyer $i$'s neighbors, and $M_i=M(C)\cap N_i$ as the subset of $i$'s neighbors who buy the information product from the seller.\footnote{The set of agent $i$'s neighbor who buy the information product $M_{i}$ depends on the contract $C$, in the following, we write $M_i=M_i(C)$ for short. } The cardinality of the sets, $m=|M(C)|$, $n=|N|$, $m_i=|M_i|$ and $n_i=|N_i|$.

Any buyer $j$ outside of $M(C)$ is not fed with the data, which is equivalent to charging a price higher than his willingness to buy, $$p_j>\frac{1}{z_0+m_jz}-\frac{1}{z_0+(m_j+1)z}.$$ To simplify the analysis, we assume $z_{j}=p_{j}=0$ for the buyer $j$ who does not buy access to the data. 
$z$ is the precision of the data collected by the seller; and $\left(p_i\right)_{i\in M(C)}$ are the prices charged for buyer $i$'s access to the data in the contract $C$. Therefore, designing an optimal contract $C$ is equivalent to choosing (i) the target $M(C)$, (ii) the precision of database $z$, and (ii) the prices $\left(p_i\right)_{i\in M(C)}$.

From Lemma \ref{lem:contract}, if the contract $C$ is optimal for the seller and the contract of any buyer $i \in M(C)$ is accepted, then the price charged for buyer $i$ must be given as follows: 
\begin{align}
p(m_i,z)\equiv\frac{1}{z_0+m_iz}-\frac{1}{z_0+(m_i+1)z}. \label{eq:price}    
\end{align}

From equation (\ref{eq:price}), it is straightforward to see that the price charged for buyer $i$ decreases as the cardinality of his neighbors in the target $m_{i}$ increases. This is intuitive since, as a buyer observes more signals from his neighbors, the marginal benefit of buying the access and generating his own signal would decrease. Therefore, the seller needs to choose between targeting a larger set of buyers with a lower price for each access to data or focusing on a small group of buyers at a higher price.

\begin{definition} \label{def:is}
The set of nodes $X$ is an \emph{independent set} of the network $G$, if any two nodes in $X$ are not connected, i.e. $\forall i, j\in X$ with $i \neq j$, there is $g_{ij}=0$.

A set of nodes is a \emph{maximal independent set} if no node can be added without violating independence. 

 An independent set of maximum cardinality is a \emph{maximum independent set}. 
 The cardinality of the maximum independent set is the independence number of the network $G$. 
\end{definition}

The independent set of a network characterizes a subset of nodes that none of the pairs of nodes is connected. The following result characterizes a condition, under which the seller would not sell the information product to any two connected buyers in the optimal contract. 

\begin{proposition}\label{prop:max} 
Let $C=(z_{i}, p_{i})_{i \in N}$ be a contract such that $M(C)$ is not an independent set of the network $G$. Then, if $z>\min_{i\in M}\frac{z_0}{1+m_i}$, $C$ is not an optimal contract for the seller. 
\end{proposition}

Proposition \ref{prop:max} provides a necessary condition for the contract $C$ targeting a non-independent set to be optimal by deriving an upper bound on the precision of the signal, beyond which the seller can strictly increases its payoff  by removing a buyer from the target $M(C)$. When the precision of the data is too high, the marginal benefit of buying access to data decreases correspondingly fast as more neighbors buy, and so selling to two neighbors becomes not a profitable option for the seller. 

In general, the seller needs to choose the precision of the data, target, and prices at the same time. The main result is presented in the following theorem.

\begin{theorem}\label{thm:main}
Assume $z_0<\frac{1}{2\sqrt{\gamma}}\frac{n+1}{2n+1}$. Then the optimal contract $C =(z_{i}, p_{i})_{i \in N}$ for network $G$ is characterized as follows:
\begin{itemize}
\item [(i)] $M(C)$ is a maximum independent set of $G$
\item [(ii)] $z=\sqrt{\frac{m}{\gamma}}-z_0$, $p_i=\frac{1}{z_0}-\sqrt{\frac{\gamma}{m}}$ \ for all \ $i\in M(C)$
\end{itemize} 
where $m=\alpha(G)$ is the independence number of network $G$.
\end{theorem}

\begin{figure}
    \hspace*{1.25cm}    
\includegraphics[scale=0.9,angle=0,origin=c]{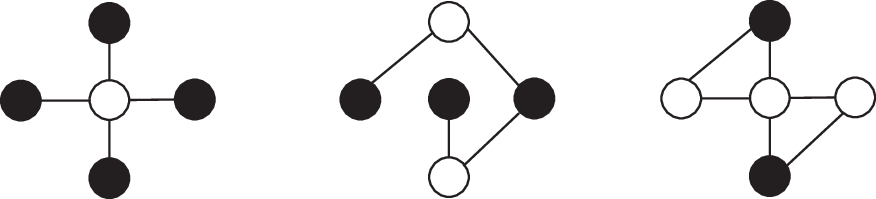}    
    \caption{Examples of $M(C)$ (shaded circles represent targeted buyers).}
    \label{fig:timing}
\end{figure}

Theorem \ref{thm:main} shows the optimal contract targets at disconnected buyers when the common prior of the buyers is noisy. Given the marginal cost of collecting data and the precision of common prior, the data collected depends only on the independence number of network $G$. As the independence number of the network increases, the precision of the data in optimal contract increases, and the price charging increases. As the precision of common prior and marginal cost of collecting data decreases, the precision and prices of the data in optimal contracts increases. In Figure \ref{fig:timing}, we provide examples of targeted buyers $M(C)$ in various networks.

\begin{remark} The bound on $z_0$ in Theorem \ref{thm:main}, $z_0<\frac{1}{2\sqrt{\gamma}}\frac{n+1}{2n+1}$, can be replaced by a uniform one $z_0<\frac{1}{4\sqrt{\gamma}}$ such that the result holds for a network $G$ of any size $n$, due to the fact that $\frac{n+1}{2n+1}$ is decreasing to $\frac{1}{2}$. 
\end{remark}

Note that a maximum independent set of a complete network consists of one node, whereas that of a star network consists of the buyers on the periphery. In these special cases, Theorem \ref{thm:main} implies the following target sets.

\begin{corollary}\label{ex:net}
  Consider the set of buyers $N=\{1, \dots, n\}$ and network $G$. 
\begin{itemize}
\item[(i)] If $G$ is a complete network, then the seller sells to only one buyer.
\item[(ii)] If $G$ is a star network with buyer $i$ being the center, then the seller sells to every buyer except for $i$.
\end{itemize}
   
\end{corollary}

\subsection{Welfare Analysis} \label{sec:welfare} 

Given the characterization of Theorem \ref{thm:main}, it is easy to check that the monopoly seller's profit $\Pi(z_{0}, G)$ in an optimal contract is given by
\begin{align*}
    \Pi\left(z_{0}, G\right)\equiv\frac{m}{z_0}+\gamma z_0-2\sqrt{\gamma m}
\end{align*}
where $m=\alpha(G)$ be the cardinality of a maximum independent set of $G$. 
\begin{proposition} \label{prop:welfare_static}
Suppose $z_0<\frac{1}{2\sqrt{\gamma}}\frac{n+1}{2n+1}$. Then, the seller's profit $\Pi(z_{0}, G)$ is strictly increasing in $m$ and strictly decreasing in $z_0$.
\end{proposition}

Notably, the seller's profit does not depend on other measurements of network structures: as long as two networks $G$ and $G'$ have the same cardinality of the maximum independent sets, the seller's profits under two networks are identical. In addition, given a fixed network $G$, the seller's profit is invariant to the target set $M(C)$ as long as it is a maximum independent set. This is in stark contrast to the buyers' surplus; the consumer surplus heavily depends on the network structure and the target set $M(C)$ even under fixed buyer network $G$. To see this, observe first that each buyer $i\in M(C)$ obtains a payoff of
\begin{align*}
    -\left(\frac{1}{z_0+z}\right)-p_i=-\frac{1}{z_0},
\end{align*}
whereas each free-rider $i\in N\setminus M(C)$ obtains a payoff of
\begin{align*}
    -\left(\frac{1}{z_0+m_i z}\right)=-\left(\frac{1}{z_0+m_i\left(\sqrt{\frac{m}{\gamma}}-z_0\right)}\right).
\end{align*}
Thus, the consumer surplus $w\left(z_0,G,M(C)\right)$ is given by
\begin{align}
    w\left(z_0,G,M(C)\right) \equiv -\sum_{i\in M(C)}\left(\frac{1}{z_0}\right)-\sum_{i\in N\setminus M(C)}\left(\frac{1}{z_0+m_i\left(\sqrt{\frac{m}{\gamma}}-z_0\right)}\right)\label{CS}
\end{align}

\begin{figure} 
    \hspace*{4cm}    
\includegraphics[scale=0.9,angle=0,origin=c]{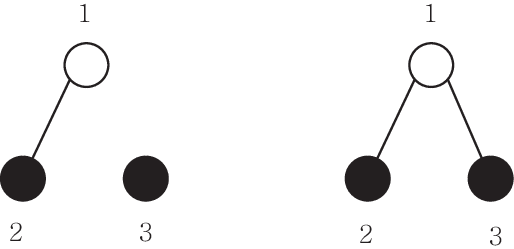}  
\caption{Different $G$'s with same $M(C)$ may induce different consumer surplus.} \label{fig:cs1}
\end{figure}

To illustrate the dependence on the network structure of consumer welfare, consider the following example of Figure \ref{fig:cs1} in which buyers $2$ and $3$ are served and the buyer $1$ free rides. 

\begin{example} \label{ex:diff_G}
Consider the set of buyers $N=\{1, 2, 3\}$ and two networks $G$ and $G'$. The only difference is that buyer $1$ is connected only to buyer $2$ in the first network ($G$), but connected to both buyers $2$ and $3$ in the second network ($G'$). $g_{12}=g'_{12}=g'_{13}=1$. 
In both cases, for the optimal contract of the seller, we have $M(C)=\{2,3\}$, which is the maximum independent set of $G$ and $G'$. The contracts in both cases are the same, with the precision of data collecting $z=\sqrt{\frac{2}{\gamma}}-z_0$ and price charging $p_{2}=p_{3}=\frac{1}{z_0}-\sqrt{\frac{\gamma}{2}}$. Thus, the profit of the seller is equal in both networks, $\Pi(z_{0}, G)=\Pi(z_{0}, G')$. 

Using equation (\ref{CS}), we can check the consumer surplus (under $z_0<\frac{1}{2\sqrt{\gamma}}\frac{n+1}{2n+1}$ so that our results apply): 
$$w(z_0,G,M(C))=-\frac{2}{z_0}-\frac{1}{\sqrt{\frac{2}{\gamma}}}<w(z_0,G',M(C))=-\frac{2}{z_0}-\frac{1}{2\sqrt{\frac{2}{\gamma}}-z_0}.$$
\end{example}

\begin{figure} 
    \hspace*{2.8cm}    
\includegraphics[scale=0.9,angle=0,origin=c]{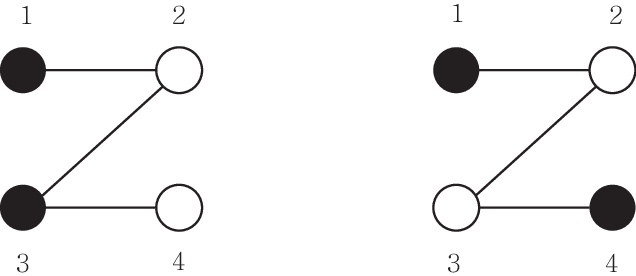}  
\caption{Different $M(C)$'s under the same $G$ may induce different consumer surplus.} \label{fig:cs2}
\label{fig:mis}
\end{figure}

The next example of Figure \ref{fig:cs2} illustrates that consumer welfare depends on the selection of target set, that is, different $M(C)$'s under a fixed network may induce different consumer surplus.

\begin{example} \label{ex:diff_M}
Consider the set of buyers $N=\{1, 2, 3, 4\}$ and network $G$ in Figure \ref{fig:cs2}. 
It is immediate to check that a maximum independent set is either $\{1,3\}$, $\{1,4\}$, or $\{2,4\}$. In the first case, buyers $1$ and $3$ are served ($M(C)=\{1,3\}$), whereas in the second case, buyers $1$ and $4$ are served ($M(C')=\{1,4\}$). The seller maximizes its profit $\Pi(z_{0}, G)$ in both cases since these are the maximum independent sets. The precision of collecting data is $z=z'=\sqrt{\frac{2}{\gamma}}-z_0$, $p_1=p_{3}=p'_{1}=p'_{4}=\frac{1}{z_0}-\sqrt{\frac{\gamma}{2}}$. 
Again, we can use equation (\ref{CS}) to calculate the consumer welfare (under $z_0<\frac{1}{2\sqrt{\gamma}}\frac{n+1}{2n+1}$ so that our results apply)
$$w(z_0,G,M(C))=-\frac{2}{z_0}-\frac{1}{2\sqrt{\frac{2}{\gamma}}-z_0}-\frac{1}{\sqrt{\frac{2}{\gamma}}}>w(z_0,G,M(C'))=-\frac{2}{z_0}-\frac{2}{\sqrt{\frac{2}{\gamma}}}.$$
The consumer welfare of contract $C$ is larger than contract $C'$, because buyer $2$ and $3$ free rides the signal from buyer $1$ and $4$ respectively for contract $C'$, while for contract $C$, buyer $2$ free rides $2$ signals and buyer $4$ free rides $1$ signal.  
\end{example}

From Example \ref{ex:diff_G} and Example \ref{ex:diff_M}, we find the consumer surplus depends on the network $G$ and the target $M(C)$ of the optimal contract $C$. In particular, the consumer surplus gets larger if more buyers observe more signals from their neighbors (who buy).  

Let $k(C)$ denote the vector of the number of links $m_i$ of free-riders with the buyers of target $M(C)$,
\begin{align*}
    k(C)\equiv (m_{1},m_{2},\cdots,m_{n-m}),
\end{align*}
where we relabel indices so that $m_1\geq m_2\geq \cdots\geq m_{n-m}.$ Note also that $\frac{k(C)}{\sum_{i=1}^{n-m}m_i}$ is a distribution. The following results are immediate from the payoff function of free-riders increasing and concave in $m_i$.\footnote{The observation regarding Figure \ref{fig:cs2} is a direct application of Proposition \ref{prop:opt_tar} (i). A further characterization of $M$ maximizing consumer welfare, however, seems challenging. The main difficulty comes from cases where two forces (i) and (ii) are in opposite directions (e.g., if most, but not all, coordinates in $k(C)$ are higher than $k(C')$ but $M(C)$ induces more symmetric allocation than $M(C')$); which effect dominates another heavily relies on the network of buyers and resulting $k(C)$'s.}

\begin{proposition}  \label{prop:opt_tar}
Let $M(C)$ and $M(C')$ be the target sets of optimal contract $C$ and $C'$. Then the consumer welfare $w(z_0,G,M(C))\geq w(z_0,G,M(C'))$ if either of the following holds:\\
(i) $k(C)\geq k(C')$.\\
(ii) $\sum_{i=1}^{n-m}m_i=\sum_{i=1}^{n-m}m_i'$ and $k(C)$ dominates $k(C')$ in the second-order stochastic dominance sense.
\end{proposition}

\subsection{Socially Efficient Contract} \label{sec:sec}

In the following, we turn to the socially efficient contract. Since the profit of the seller is $\pi(C)=\sum_{i\in N}p_i-c(z_1,...,z_n)$, and the buyer $i$'s payoff is $\pi_{i}(a_{i}, C) =-\left(\frac{1}{z_0+\sum_{j=1}^{n} g_{i j} z_{j}}\right)-p_i$, we have the social welfare function as follows:
$$SW(C) = \pi(C) + \sum_{i \in N} \pi_{i}(a_{i}, C) =-\left(\frac{1}{z_0+\sum_{j=1}^{n} g_{i j} z_{j}}\right) -\gamma z.$$
Note that the socially efficient contract must involve $M(C)=N$: since the total cost of providing data does not depend on the number of served buyers, it is always welfare-improving to provide access to the data to additional buyers.

\begin{proposition}\label{prop:welfare}
The socially efficient contract $C^*=(z^*, p_{i})_{i \in N}$ is characterized as follows: 
\begin{itemize}
    \item [(i)] $z_{i}=z^* > 0$ for any $i \in N$. 
    \item [(ii)] $z^*$ solves the following equation:
    \begin{align*}
        \sum_{i\in N}\left(\frac{n_i+1}{\left(z_0+(n_i+1)z^* \right)^2}\right)=\gamma.
    \end{align*}
\end{itemize}
\end{proposition}

In general, there is no closed-form solution for the socially efficient precision $z^*$. To make further progress, we thus consider specific networks allowing for explicit derivation of $z^*$ and compare it with the seller-optimal $z^{o}$ in the following example. 

\begin{example} \label{ex:precision}
Consider the set of buyers $N=\{1, \dots, n\}$, network $G$ is a regular network with every node with degree $d$.  
\begin{itemize}
    \item [(i)] Consider the network of a cycle $G$ with $d=2$ and the independence number $m=\left[\frac{n}{2}\right]$ . The socially efficient level of data quality is $z^* = \sqrt{\frac{n}{3 \gamma}}-\frac{z_{0}}{3}$, and the data quality in an optimal contract is $z^{o}=\sqrt{\frac{m}{\gamma}}-z_{0}$. For $z_{0} < \frac{3}{2\sqrt{\gamma}} \left(\sqrt{m}-\sqrt{\frac{n}{3}}\right)$, we have $z^* < z^{o}$. 
\item [(ii)] Consider a complete network $G$ with $d=n-1$ and the independence number $m=1$. The socially efficient level of data quality is $z^* = \sqrt{\frac{1}{\gamma}}-\frac{z_{0}}{n}$, and the data quality in an optimal contract is $z^{o}=\sqrt{\frac{1}{\gamma}}-z_{0}$. For any $z_{0}$, we have $z^* > z^{o}$. 
\end{itemize}

\end{example}

Consider a connected graph $G$. 
From Example \ref{ex:precision}, the socially efficient precision is smaller than the precision of the optimal contract if the network is not complete. The intuition behind the example is the following: for the socially efficient contract, the information has a spillover effect, so lower precision of data is needed. On the other hand, the seller tries to maximize the profit, and it has an incentive to sell data with higher precision to the targeted buyers because their willingness to pay for the only information source is higher because they are disconnected. In an optimal contract, therefore, the seller sells higher precision of data than in the socially efficient level. The following theorem summarizes this observation.

\begin{theorem} \label{thm:eff_prec}
As the precision of common prior $z_{0}$ approaches $0$, the precision $z^*$ of a socially efficient contract is smaller than the precision $z^{o}$ of an optimal contract if the network $G$ is neither a complete graph nor a graph composed of complete subgraphs.

\end{theorem}

\subsection{Pareto Efficient Network} \label{sec:opt_net}
Finally, we provide a brief discussion about Pareto-efficient networks, which are defined as follows. 

\begin{definition} \label{def:pe_net}
A network $G$ is said to be \emph{Pareto-efficient} if the payoffs of the buyers cannot be improved by modifying the network to some $G'$ without decreasing the seller's profit.  
\end{definition}

Note that, without hurting the profit of the seller, consumer welfare is maximized when there exists a link between any pair of a free rider and a purchaser of data. This implies that any network with a core-periphery structure is Pareto-efficient, where a core-periphery network is defined as: for any $i \in N \setminus M(C)$, $g_{ij}=1$ for any $j \neq i$; for any $i,j \in M(C)$, $g_{ij}=0$.  

\begin{proposition} \label{prop:pe_net}
	A core-periphery network, in which there are $m$ nodes in the periphery and $n-m$ nodes in the core, is Pareto-efficient with $n$ nodes and independence number $m$. 
\end{proposition}

Given the number of nodes $n$ and independence number $m$, a Pareto-efficient network provides the most connections between the free riders and the buyers of data. The set of buyers is a maximum independent set where there is no connection between them. The graph with free riders connected with all purchasers provides the highest consumer welfare, while the connections between free riders do not affect the profit of the seller and consumer welfare. Thus, it is incentive-compatible for the seller to add the connections between free riders, keeping the set of buyers as the maximum independent set.

\section{Conclusion} \label{sec:conclusion}
In this paper, we studied a stylized model of a monopoly data seller when information-sharing network exists among data buyers. We show that, if the buyers' prior information is sufficiently noisy, the optimal selling strategy is characterized by a maximum independent set, which is the largest set of buyers who do not have information-sharing link at all. In addition, the precision of the seller's data decreases in the number of information-sharing links among buyers, but it is \emph{higher} than the socially efficient level of precision. 

There are several limitations to our paper. First of all, we have abstracted away from coordination motives by assuming that the buyers' utility  depends only on the true state of the world, not the others' actions. Although we believe that our results continue to hold as long as the coordination motive is relatively weaker, it is still important to check whether our results are robust to such extenstions. A more important question that this paper does not address, however, would be the impact of competition among data sellers.  We leave these important questions for future research.

 \bibliographystyle{aea}
 \bibliography{info}

\section*{Appendix: Proofs}

\paragraph{Proof of Proposition \ref{prop:max}.}

Let $i\in M(C)$ with a maximal $m_i$. The idea is that if $i$ is served, there are many users in $M(C)$ that free ride his information, which the seller tries to avoid such that they can charge more from those users in $M_i$. We consider removing $i$ from $M(C)$, and compare the induced cost and benefit. We will show that the benefit strictly dominates the cost hence by doing so the seller's revenue is strictly improved, which implies that $M(C)$ is not optimal provided that it is not an independent set. The cost is the reduction of revenue raised from user $i$, which is $p_i=\frac{1}{z_0+m_iz}-\frac{1}{z_0+(m_i+1)z}$. The possible benefit is the change in the price sum charged over all $j\in M_i$, which is ($p_j$ denotes the price before and $p'_j$ denotes the price after the change)

\begin{equation*}
\begin{aligned}
\sum_{j\in M_i} (p'_j-p_j)=& \ \sum_{j\in M_i} \bigg( \Big(\frac{1}{z_0+(m_j-1)z}-\frac{1}{z_0+m_jz}\Big)  - \Big(\frac{1}{z_0+m_jz}-\frac{1}{z_0+(m_j+1)z}\Big)\bigg)\\
= & \ \sum_{j\in M_i} \bigg(\frac{1}{z_0+(m_j-1)z}+\frac{1}{z_0+(m_j+1)z}-\frac{2}{z_0+m_jz}\bigg).
\end{aligned}
\end{equation*}
Let us define $\Delta(m_j):=\frac{1}{z_0+(m_j-1)z}+\frac{1}{z_0+(m_j+1)z}-\frac{2}{z_0+m_jz}$, which has the following properties:
\begin{itemize}
\item [(1)]  $\Delta(m_j)>0$ since the function $f(x)=\frac{1}{x}$ is convex on $(0, \infty)$;
\item  [(2)] $\Delta'(m_j)=\bigg(\frac{2z}{\big(z_0+m_jz\big)^2} - \bigg(\frac{z}{\big(z_0+(m_j-1)z\big)^2}+\frac{z}{\big(z_0+(m_j+1)z\big)^2}\bigg)\bigg)<0$ since the function $f(x)=\frac{1}{x^2}$ is convex.
\end{itemize}
This implies that $\Delta(m_j)$ is strictly decreasing in $m_j$ for each $j\in M_i$. Since $m_j\in [1, m_i]$ by the choice of the vertex $i$ in $M(C)$, we obtain that for all $j\in M_i$,
$$p_j'-p_j=\Delta(m_j)\geq \frac{1}{z_0+(m_i-1)z}+\frac{1}{z_0+(m_i+1)z}-\frac{2}{z_0+m_iz}$$
hence
$$\sum_{j\in M_i}(p_j'-p_j)\geq  \frac{m_i}{z_0+(m_i-1)z}+\frac{m_i}{z_0+(m_i+1)z}-\frac{2m_i}{z_0+m_iz}$$
In the end, it remains to prove that 
$$\frac{m_i}{z_0+(m_i-1)z}+\frac{m_i}{z_0+(m_i+1)z}-\frac{2m_i}{z_0+m_iz}> p_i,$$
which is equivalent to 
\begin{align*}
  &\frac{m_i}{z_0+(m_i-1)z}+\frac{m_i+1}{z_0+(m_i+1)z}> \frac{2m_i+1}{z_0+m_iz}\\
  \iff&\left(\frac{m_i}{2m_i+1}\right)\left(\frac{1}{z_0+(m_i-1)z}\right)+\left(\frac{m_i+1}{2m_i+1}\right)\left(\frac{1}{z_0+(m_i+1)z}\right)>\frac{1}{z_0+m_iz}
\end{align*}
A tedious computation shows that this is equivalent to
\begin{align*}
    \frac{z((m_i+1) z-z_0)}{(1+2m_i)(z_0+(m_i-1)z)(m_iz+z_0)((m_i+1)z+z_0)}>0
    \iff z>\frac{z_0}{1+m_i}
\end{align*} \qedsymbol

\bigskip

\paragraph{Proof of Theorem \ref{thm:main}.}
Once the optimal $M(C)$ is a maximum independent set, the condition for $z$ and $p_i$ are easily derived.\footnote{$\pi(C)=m\big(\frac{1}{z_0}-\frac{1}{z_0+z}\big)-\gamma z$, so the FOC gives us the optimal $z=\sqrt{\frac{m}{\gamma}}-z_0$ and $p_i=\frac{1}{z_0}-\sqrt{\frac{\gamma}{m}}$ for all $i\in M(C)$.} In the following, we prove that the optimal contract $C=(z_i,p_i)_{i\in N}$ should target an independent set $M(C)$ of the network $G$, based on which we find by calculation\footnote{Let $m'\leq m$ be the size of a targeted independent set. The optimal profit is $\pi(C)=m'\big(\frac{1}{z_0}-\sqrt{\frac{\gamma}{m'}}\big)-\gamma\big(\sqrt{\frac{m'}{\gamma}}-z_0\big)=\frac{m'}{z_0}-2\sqrt{m'\gamma}+\gamma z_0$, which is increasing in $\sqrt{m'}$ since the minimizer of the quadratic function of $\sqrt{m'}$ is $z_0\sqrt{\gamma}<1$ and $\sqrt{m'}$ is defined on $[1, \infty)$. } that a maximum independent set is the optimal solution. 

Suppose that a current $M(C)$ is not an independent set of $G$, and $C=(z_i,p_i)_{i\in N}$ is optimal under $M(C)$ (necessary condition for $C$ to be optimal for $G$). $M(C)$ not being an independent set implies that there are at least two nodes, $i, j \in M(C)$, that are connected. Denote by $i^*$ with maximum $m_{i^*}$. According to Proposition \ref{prop:max}, the sufficient condition to $C$ not to be optimal should fail, i.e. 
\begin{equation}\label{eq:fail}
\begin{aligned}
z<\frac{z_0}{1+m_{i^*}}.
\end{aligned}
\end{equation}
We are using this condition to derive a contradiction to the assumption that $C$ is optimal, in particular, we show that the seller can increase $z$ to get strictly higher profit. To do so, we prove that the marginal revenue net of the marginal cost in $z$ is strictly positive. Indeed, $$\pi(C)=\sum_{i\in M(C)} \big(\frac{1}{z_0+m_iz}-\frac{1}{z_0+(m_i+1)z}\big)-\gamma z,$$
so it is sufficient for us to show that
\begin{equation}\label{eq:MR}
\begin{aligned}
MR_i(C)=\frac{m_i+1}{(z_0+(m_i+1)z)^2}-\frac{m_i}{(z_0+m_iz)^2}>\frac{\gamma}{m}, \text{ \ for all \ } i\in M(C). 
\end{aligned}
\end{equation}
Notice that 
$$MR_i(C)=\frac{z_0^2-m_i(m_i+1)z^2}{(z_0+(m_i+1)z)^2(z_0+m_iz)^2}:=\beta(z),$$ and it remains to show that 

$$\beta(z)=\frac{z_0^2-m_i(m_i+1)z^2}{(z_0+(m_i+1)z)^2(z_0+m_iz)^2}>\frac{\gamma}{m}, \text{ \ for all \ } i\in M(C).$$

We notice that the function $\beta(z)$ is decreasing in $z$ on $[0, \infty)$. Now we use the equation (\ref{eq:fail}) implied the assumption that $C$ being optimal the show that $\beta(z)>\frac{\gamma}{m}$ for all $z<\frac{z_0}{1+m_{i^*}}$. Since $i^*$ is the user in $M(C)$ with the maximal $m_i$, we should have $z<\frac{z_0}{1+m_{i}}$ for all $i\in M(C)$. It is sufficient for us to check the inequality $\beta(z)>\frac{\gamma}{m}$ for $z=\frac{z_0}{1+m_{i}}$ for all $i$. Indeed, we have 

$$\beta\left(\frac{z_0}{1+m_{i}}\right) =\frac{z_0^2\left(1-\frac{m_i}{m_i+1}\right)}{z_0^4\left(1+1\right)^2\left(1+\frac{m_i}{m_i+1}\right)^2}=\frac{1}{4\left(m_i+1\right)z_0^2\left(1+\frac{m_i}{m_i+1}\right)^2}.
$$
To show that $\beta\big(\frac{z_0}{1+m_{i}}\big)>\frac{\gamma}{m}$, we use the facts $m_i+1\leq m$, $\frac{m_i}{m_i+1}\leq \frac{n}{n+1}$ and that $z_0<\frac{1}{2\sqrt{\gamma}}\frac{n+1}{2n+1}$. 
This completes the proof. \qedsymbol

\paragraph{Proof of Proposition \ref{prop:welfare_static}.}
From Theorem \ref{thm:main}, the seller's profit of the optimal contract is $\Pi(z_{0}, G) = \frac{m}{z_0}+\gamma z_0-2\sqrt{\gamma m}$.  For $z_0<\frac{1}{2\sqrt{\gamma}}\frac{n+1}{2n+1}$, $\frac{\partial \Pi(z_{0}, G)}{\partial m}=\frac{1}{z_{0}}-\sqrt{\frac{\gamma}{m}} > (\frac{2(2n+1)}{n+1}-\sqrt{\frac{1}{m}})\sqrt{\gamma}$.  Since $m \geq 1$ and $\frac{2(2n+1)}{n+1} \geq 2$, we have $\frac{\partial \Pi(z_{0}, G)}{\partial m} > 0$.  $\frac{\partial \Pi(z_{0}, G)}{\partial z_{0}}=-\frac{m}{z_{0}^{2}} + \gamma < (- (\frac{2(2n+1)}{n+1})^{2}m + 1)\gamma$. Since $m \geq 1$, $(\frac{2(2n+1)}{n+1})^{2} \geq 4$, we have $\frac{\partial \Pi(z_{0}, G)}{\partial z_{0}} < -3\gamma < 0$.   
Therefore, the seller's profit of the optimal contract is strictly increasing in $m$ and strictly decreasing in $z_0$. \qedsymbol

\paragraph{Proof of Proposition \ref{prop:opt_tar}.}
From Equation (\ref{CS}), the consumer welfare is $$w\left(z_0,G,M(C)\right) = -\sum_{i\in M(C)}\left(\frac{1}{z_0}\right)-\sum_{i\in N\setminus M(C)}\left(\frac{1}{z_0+m_i\left(\sqrt{\frac{m}{\gamma}}-z_0\right)}\right).$$ 

For optimal contracts $C$ and $C'$, the first component of the consumer welfare are the same, $-\sum_{i\in M(C)}\left(\frac{1}{z_0}\right) = -\sum_{i\in M(C')}\left(\frac{1}{z_0}\right) = - \frac{m}{z_0}$. The second component of the consumer welfare are $-\sum_{i=1}^{n-m}\frac{1}{z_0+m_i\left(\sqrt{\frac{m}{\gamma}}-z_0\right)}$ and $-\sum_{i=1}^{n-m}\frac{1}{z_0+m_i'\left(\sqrt{\frac{m}{\gamma}}-z_0\right)}$. 
	
(i) $k(C)\geq k(C')$, we have $\frac{1}{z_0+m_i\left(\sqrt{\frac{m}{\gamma}}-z_0\right)} \leq \frac{1}{z_0+m_i'\left(\sqrt{\frac{m}{\gamma}}-z_0\right)}$ for any $1 \leq i \leq n-m$. Thus, $-\sum_{i=1}^{n-m}\frac{1}{z_0+m_i\left(\sqrt{\frac{m}{\gamma}}-z_0\right)} \geq -\sum_{i=1}^{n-m}\frac{1}{z_0+m_i'\left(\sqrt{\frac{m}{\gamma}}-z_0\right)}$. The consumer welfare satisfies $w(z_0,G,M(C))\geq w(z_0,G,M(C'))$. 

(ii) Consider the case $n-m=2$, such that $m_{1}'< m_{1} \leq m_{2} < m_{2}'$ and $m_{1} + m_{2} = m_{1}' + m_{2}'$. Then we have $\frac{1}{z_0+m_1'\left(\sqrt{\frac{m}{\gamma}}-z_0\right)} + \frac{1}{z_0+m_2'\left(\sqrt{\frac{m}{\gamma}}-z_0\right)} >\frac{1}{z_0+m_1\left(\sqrt{\frac{m}{\gamma}}-z_0\right)} + \frac{1}{z_0+m_2\left(\sqrt{\frac{m}{\gamma}}-z_0\right)}$. Following this analogy, $\frac{k(C')}{\sum_{i=1}^{n-m}m_i'}$ is a mean preserve spread of $\frac{k(C)}{\sum_{i=1}^{n-m}m_i}$, the consumer welfare satisfies $w(z_0,G,M(C))> w(z_0,G,M(C'))$. \qedsymbol

\paragraph{Proof of Proposition \ref{prop:welfare}.}
Given the contract $C=(z_i,p_i)_{i\in N}$, the social welfare is given as
\begin{align*}
    &\sum_{i\in M(C)}p_i-\gamma z+\sum_{i\in M(C)}\left(-\frac{1}{z_0+(m_i+1)z}-p_i\right)+\sum_{i\in N\setminus M(C)}\left(-\frac{1}{z_0+m_i z}\right)\\
    =&\sum_{i\in M(C)}\left(-\frac{1}{z_0+(m_i+1)z}\right)+\sum_{i\in N\setminus M(C)}\left(-\frac{1}{z_0+m_i z}\right)-\gamma z 
\end{align*}
Clearly, we have
\begin{align*}
    &\sum_{i\in M(C)}\left(-\frac{1}{z_0+(m_i+1)z}\right)+\sum_{i\in N\setminus M(C)}\left(-\frac{1}{z_0+m_i z}\right)-\gamma z\\
    \leq&\sum_{i\in N}\left(-\frac{1}{z_0+(n_i+1)z}\right)-\gamma z
\end{align*}
This shows that we have $M(C)=N$ in a socially optimal contract. 

When  $M(C)=N$, the social welfare is $\sum_{i\in N}\left(-\frac{1}{z_0+(n_i+1)z}\right)-\gamma z$. Now, the condition for optimal $z$ in the statement follows from the first-order condition to maximize the social welfare. Therefore, the socially efficient precision $z^*$ satisfies $\sum_{i\in N}\left(\frac{n_i+1}{\left(z_0+(n_i+1)z^* \right)^2}\right)=\gamma$. 
\qedsymbol

\paragraph{Proof of Theorem \ref{thm:eff_prec}.}

The precision $z^{o}$ of an optimal contract satisfies $z^{o}=\sqrt{\frac{m}{\gamma}}-z_0$, as $z_{0}$ approaches $0$, the limit is $z^{o}_{lim}= \lim_{z_{0} \rightarrow 0} \sqrt{\frac{m}{\gamma}}-z_0 =\sqrt{\frac{m}{\gamma}}$, $m$ is the independence number. 

From Proposition \ref{prop:welfare}, the socially efficient level of data collection $z^*$ solves the equation 
$\sum_{i\in N}\left(\frac{n_i+1}{\left(z_0+(n_i+1)z^* \right)^2}\right)=\gamma$. 
As the precision of common prior $z_{0}$ approaches $0$, take the limit of $z_{0}$, we have $\lim_{z_{0} \rightarrow 0}\sum_{i\in N}\left(\frac{n_i+1}{\left(z_0+(n_i+1)z^* \right)^2}\right) = \sum_{i\in N}\frac{1}{(n_i+1)z^{*2}}$. Thus, the socially efficient level of data collection $z^*$ is approaching the limit $z_{lim}^{*} = \sqrt{\sum_{i\in N}\frac{1}{(n_i+1)\gamma}}$. 

To compare the limit of precision of collected information for seller's optimal contract $z^{o}_{lim}$ and socially efficient contract $z_{lim}^{*}$, it is equivalent to compare the independence number $m$ and the summation of the inverse of the degrees of nodes plus one $\sum_{i\in N}\frac{1}{(n_i+1)}$. The Caro-Wei bound of the independence number shows $\sum_{i\in N}\frac{1}{(n_i+1)} \leq m$. Based on the proof of the probabilistic method in \cite{alon2016probabilistic}, we show the $\sum_{i\in N}\frac{1}{(n_i+1)} = m$ if and only if the network $G$ is either a complete graph or a graph composed of complete subgraphs.

The set of the agents is $N = \{1,\dots,n\}$, and assume set of permutations of $N$ is $S_{N}$. Let $\tau: N \mapsto N$ be a permutation of $N$, $\tau$ is a bijection. We define a map $A: S_{N} \mapsto 2^{N}$, $A(\tau) =\{i \in N | \tau(i) < \tau(j), \forall j \neq i \in N, g_{ij} = 1\}$. 
Since $i_{1} \in A(\tau)$ for $\tau(i_{1})=1$, for any permutation $\tau$, $A(\tau)$ is not empty,  
If $i \in A(\tau)$, then none of node $i$'s neighbor is not in the set $A(\tau)$. For any node $j$ with $g_{ij} = 1$, $i \in A(\tau)$, from the definition of $A$, we have $\tau(i) < \tau(j)$, then $j$ has a neighbor $i$, such that $\tau(j) > \tau(i)$, thus, $j \notin A(\tau)$. 
Therefore, $A(\tau)$ is an independent set of the network $G$ for any $\tau \in S_{N}$. The independence number $m$ is the cardinality of the maximum independent set, $|A(\tau)| \leq m$ for any $\tau$. Assume the permutations are uniformly distributed, the probability of any $\tau$ is $\text{Prob}(\tau)=\frac{1}{n!}$. $i$ has $n_{i}$ neighbors, $i \in A(\tau)$ if and only if the $\tau(i) = \min_{j\in N_{i}\cup \{i\}}\tau(j)$. 


$|N_{i}\cup \{i\}|=n_{i}+1$, there are $(n_{i}+1)!$ permutations of $N_{i}\cup \{i\}$ and $n_{i}!$ permutations of $N_{i}$. So there are $\tbinom{n}{n_{i}+1}n_{i}!(n-n_{i}-1)!$ permutations of $N$, such that $i \in A(\tau)$. 
Thus, the probability of any node $i$ is in the set $A(\tau)$ is $\text{Prob}(i \in A(\tau))=\frac{\tbinom{n}{n_{i}+1}n_{i}!(n-n_{i}-1)!}{n!}=\frac{1}{n_{i}+1}$. 
The expectation of set $A(\tau)$ is $E(|A(\tau)|)= \sum_{i \in N}\text{Prob}(i \in A(\tau)) = \sum_{i \in N}\frac{1}{n_{i}+1}$. Since $|A(\tau)| \leq m$ for any $\tau$, $E(|A(\tau)|) \leq m$, we have $\sum_{i\in N} \frac{1}{n_{i}+1} \leq m$.

In the proof of the Caro-Wei bound above, $\sum_{i\in N} \frac{1}{n_{i}+1} = m$ if and only if for any $\tau$, $|A(\tau)| = m$. 

If the network $G$ is connected, we show that there exists a permutation $\tau_{0}$, such that $|A(\tau_{0})| = 1$. For any node $i$, $i$'s neighbors are $N_{i}$, assume the distance between node $i$ and $j$ is $\ell(i,j)$, $\ell_{i} = \max_{j \in N}\ell(i,j)$.
Let $A_{0} = \{i\}$, $A_{k} = A_{k-1} \cup (\bigcup_{j \in A_{k-1}}N_{j})$ for $1 \leq k \leq \ell_{i}$. Since $\ell(i,j) \leq \ell_{i}$, $A_{\ell_{i}} = N$.  
We find a permutation $\tau_{0}$, such that, $\tau_{0}(i)=1$, $|A_{k-1}| + 1 \leq \tau_{0}(j) \leq |A_{k}|$ for any $j \in A_{k} \setminus A_{k-1}$, $1 \leq k \leq \ell_{i}$. Then we show $A(\tau_{0}) =\{i\}$. For any $j \in A_{k} \setminus A_{k-1}$, $1 \leq k \leq \ell_{i}$, from the construction of $A_{k}$, we have $j \in \bigcup_{j' \in A_{k-1}}N_{j'}$, so $\tau_{0}(j') \leq |A_{k-1}| < |A_{k-1}|+1 \leq \tau_{0}(j)$. There exists $j' \in A_{k-1}$, $\tau_{0}(j')<\tau_{0}(j)$ with $g_{j'j}=1$, so $j \notin A(\tau_{0})$. 
Thus, $A(\tau_{0})=\{i\}$. Since $|A(\tau_{0})|=1$, for any $m > 1$, we have $E(|A(\tau)|) < m$. 
$E|A(\tau)| = m$ if and only if $m=1$. Moreover, $m=1$ if and only if the connected network is a complete graph. 

Following the analogy, if the network $G$ is not connected, we apply the analysis above to each components of $G$, and show that the equation $\sum_{i\in N}\frac{1}{n_i+1} = m$ holds if and only if every component of network $G$ is a complete subgraph. 

Therefore, if the network $G$ is neither a complete graph, nor a network composed of complete subgraphs, we have $m > \sum_{i\in N}\frac{1}{n_i+1}$.  
Thus, in this case, when $z_{0}$ is small enough, we have $z^{*} < z^{o}$. 
\qedsymbol

\paragraph{Proof of Proposition \ref{prop:pe_net}.}

Consider a core-periphery network $G$, assume the set of buyers of the periphery is $G_{P}$ and the set of buyers of the core is $G_{C}$. The buyer $i$ of $G_{C}$ is connected with any other buyer, and the buyer $j$ of $G_{P}$ is only connected to the buyers of $G_{C}$. For any $i \in G_{C}$, if $i \in M(C)$, then $M(C) =\{i\}$. Thus, the maximum independent set is $G_{P}$, $m = |G_{P}|$. 

For the case of maximum independent set of $m$ agents, the profit of the optimal contract of the seller is $\Pi(z_{0}, G)= \frac{m}{z_0}+\gamma z_0-2\sqrt{\gamma m}$. For the core-periphery network $G$, the expected utility of any buyer $i \in G_{C}$ is $-\frac{1}{z_0+m\left(\sqrt{\frac{m}{\gamma}}-z_0\right)}$, and the expected of any buyer $j \in G_{P}$ is $-\frac{1}{z_0}$. 


Suppose the network $G'$ is a Pareto improvement over $G$, then either the consumer welfare or the profit of the optimal contract of the seller is improved without hurting the other. Assume $m'=\alpha(G')$, from Proposition \ref{prop:welfare_static}, the profit of the optimal contract of the seller satisfies $\Pi(z_{0}, G') \geq \Pi(z_{0}, G)$ if and only if $m' \geq m$. Assume the optimal contract for network $G'$ is $C'$. 

If $m' = m$, then the expected utility of buyer $j$ who buys the data is $-\frac{1}{z_0}$, and the expected utility of buyer $i$ who does not buy the data is $-\frac{1}{z_0+m_{i}\left(\sqrt{\frac{m'}{\gamma}}-z_0\right)}$. The neighbors of buyer $i$ who buy the data is a subset of the target $M(C')$, so $m_{i} \leq m'$. Thus, $-\frac{1}{z_0+m_{i}\left(\sqrt{\frac{m'}{\gamma}}-z_0\right)} \leq -\frac{1}{z_0+m\left(\sqrt{\frac{m}{\gamma}}-z_0\right)}$. Thus, none of the buyers and the seller gets strictly better off for network $G'$. 


If $m' > m$, then the seller get strictly better off in the optimal contract. The expected utility of buyer $i$ of $M(C')$ is $-\frac{1}{z_0}$, there exists at least one buyer $i \in M(C')$ who does not buy the data in network $G$. We have $-\frac{1}{z_0+m\left(\sqrt{\frac{m}{\gamma}}-z_0\right)} > -\frac{1}{z_0}$, so the buyer $i$ gets worse off for the network $G'$. It is a contradiction for the network $G'$ to be a Pareto improvement over $G$. 
\qedsymbol

\end{document}